\documentstyle[12pt,epsf]{article}

\begin{document}

%
%
\begin{titlepage}

\begin{flushright}
OU-HET 282 \\
hep-th/9711117 \\
November 1997
\end{flushright}
\bigskip
\bigskip

\begin{center}
{\Large \bf
A Proof of Brane Creation
via $M$-theory
}

\bigskip
Toshio Nakatsu, Kazutoshi Ohta, Takashi Yokono and Yuhsuke Yoshida\\
\bigskip
{\small \it
Department of Physics,\\
Graduate School of Science, Osaka University,\\
Toyonaka, Osaka 560, JAPAN
}
\end{center}
\bigskip
\bigskip
\begin{abstract}
We study configurations of a single $M$ fivebrane
in the geometry ${\bf R}^7 \times Q$
where $Q$ is the Taub-NUT space.
Taking the IIA limit at each value of their modulus,
two possibilities of the IIA configuration are
revealed. One consists of
a NS fivebrane and a D sixbrane while the other
consists of a NS fivebrane, a D sixbrane and a D fourbrane.
In the latter case the fourbrane is shown to be suspended
by the fivebrane and the sixbrane.
This appearance of fourbrane can be interpreted
in Type IIA picture as the result of the crossing of
the fivebrane and the sixbrane.

\end{abstract}

\end{titlepage}

         The possibility of brane creation in string theory
was first pointed out in \cite{H-W}.
Several proofs for it have been given \cite{brane creation 1},
\cite{brane creation 2},\cite{brane creation 3},
\cite{brane creation 4}
mainly based on the so-called anomaly-inflow argument \cite{Harvey}
and M(atrix) theory description \cite{matrix}.
In this paper
we would like to add another proof of brane creation
based on a different perspective.
Namely we examine configurations of a M fivebrane in the
eleven-dimensional geometry ${\bf R}^7 \times Q$
where $Q$ is the Taub-NUT space.
This eleven-dimensional background is
the classical solution of eleven-dimensional SUGRA which
provides \cite{Townsend}
a single D sixbrane solution of IIA string theory.
In these configurations
two-dimensional part of the worldvolume of fivebrane
is embedded holomorphically into $Q$.
With a detailed study on its holomorphic embedding
we consider its IIA limit at each value of the modulus of the
configurations. It is shown that there are two possibilities
of the IIA configuration, one of which is the configuration
consisting of a NS fivebrane and a D sixbrane while the other
is the configuration consisting of a NS fivebrane, a D sixbrane
and a D fourbrane. In the latter case the fourbrane is actually
suspended between the fivebrane and the sixbrane.
This appearance of fourbrane can be
interpreted as the result of the crossing of the fivebrane
and the sixbrane in IIA picture.

~

~

    Consider a single D sixbrane with
worldvolume  $(x^0,x^1,x^2,x^3,x^7,x^8,x^9)$.
D sixbrane is a magnetic source of the $U(1)$
gauge potential $A$ which is a bosonic ingredient
of IIA SUGRA multiplet.
Notice that, compactifying the eleventh dimension into a circle,
the eleven-dimensional metric tensor $G_{MN}$
($0 \leq M,N \leq 10$), which is a bosonic ingredient
of eleven-dimensional SUGRA multiplet, gives rise to the
$U(1)$ gauge potential $A$.
Therefore D sixbrane is the Kaluza-Klein monopole.
Though the four-dimensional space $(x^4,x^5,x^6,x^{10})$
transversal to the sixbrane is originally a flat space
${\bf R}^3 \times S^1$ $(\equiv Q_0)$,
the existence of sixbrane makes it a curved one.
It actually becomes \cite{Townsend} the Taub-NUT space $Q$.
To describe the Taub-NUT geometry let us introduce the coordinates
$(\vec{r},\sigma)$
\begin{eqnarray}
\vec{r} \equiv (2x^4/R,2x^5/R,2x^6/R)~~,~~
\sigma \equiv 2x^{10}/R,
\end{eqnarray}
where $R$ is the radius of $S^1$. The Taub-NUT metric acquires the
standard form \cite{Gibbons-Hawking}
\begin{eqnarray}
ds^2=
\frac{V}{4}d \vec{r}^2
+\frac{1}{4V}(d \sigma +
     \vec{\omega}\cdot d \vec{r})^2,
\label{TN metric}
\end{eqnarray}
where
\begin{eqnarray}
V=1+\frac{1}{|\vec{r}-\vec{r}_0|},
\label{V}
\end{eqnarray}
$\vec{r}_0$ denotes the position of the sixbrane.
And $\vec{\omega}$ is chosen so that it satisfies the relation
\begin{eqnarray}
\vec{\nabla} \times \vec{\omega} = \vec{\nabla} V.
\label{omega}
\end{eqnarray}
Notice that the $U(1)$ gauge potential $A$ can be read
from eq.(\ref{TN metric}) as $A$ $=\vec{\omega}\cdot d \vec{r}$.
Since $\vec{\omega}$ is a quantity determined by relation (\ref{omega}),
it has an ambiguity depending on the direction of the Dirac string
associated with the sixbrane.
It turns out useful to fix the gauge potential including the Dirac string.
By separating the three-vector $\vec{r}$ into two parts
$\vec{r}=(v ,b)$ where
\begin{eqnarray}
v \equiv \frac{2(x^4+ix^5)}{R}~~,~b \equiv \frac{2x^6}{R} ,
\label{v and b}
\end{eqnarray}
the $U(1)$ gauge potential $A$ can be put into the form
\begin{eqnarray}
A = \mbox{Im} \left\{ \delta~ d v \right\},
\label{A}
\end{eqnarray}
where we introduce $\delta$ as
\begin{eqnarray}
\delta \equiv
\frac{b+b_0+\sqrt{(b+b_0)^2+|v+e_0|^2}}{\sqrt{(b+b_0)^2+|v+e_0|^2}}
\frac{1}{|v+e_0|}
\label{lambda}
\end{eqnarray}
Notice that the position of the sixbrane is denoted by
$\vec{r}_0=(-e_0,-b_0)$ in eq. (\ref{lambda}).
With the above choice of the gauge potential
the field strength $F=dA$ acquires the Dirac string
of the following type
\begin{eqnarray}
F \sim \pi i \theta(b+b_0) \delta^{(2)}(v+e_0) dv \wedge d \bar{v}.
\label{Dirac string}
\end{eqnarray}
The Dirac string is now realized as the semi-infinite line
in the $(v,b)$-space which runs from $(-b_0,-e_0)$,
parallel with the $b$-axis, into $(+\infty,-e_0)$.
What is a counterpart of the Dirac string in the Taub-NUT geometry?
Recall that the Taub-NUT space can be regarded as the
fibered space which base and fiber are respectively
the $(v,b)$-space and $S^1$.
There exists a singular fiber at $(v,b)=(-e_0,-b_0)$.
This is the so-called NUT singularity.
With this understanding of the Taub-NUT space,
an inverse image of the Dirac string in the Taub-NUT space
gives its counterpart.
It is the non-compact two-cycle
which is precisely projected onto the Dirac string.

               The Taub-NUT space $Q$ admits to have an hyper-K\"ahler
structure.
Especially it is possible to choose a complex structure so that the Taub-NUT
metric
(\ref{TN metric}) becomes K\"ahler.
It is given \cite{NOYY} by introducing the following
holomorphic coordinates $(v,y)$
\begin{eqnarray}
v &=&
\frac{2(x^4+ix^5)}{R} ,
\label{v} \\
y &\equiv&
e^{- \frac{b+i \sigma}{2}}
\left( -b-b_0+
   \sqrt{(b+b_0)^2+|v+e_0|^2} \right)^{\frac{1}{2}} .
\label{y}
\end{eqnarray}
With these  holomorphic coordinates the Taub-NUT metric (\ref{TN metric})
\footnote{$A=\vec{\omega}\cdot d \vec{r}$
has the form given in eq. (\ref{A}).}
can be rewritten as
\begin{eqnarray}
ds^2=
\frac{V}{4}dv d \bar{v}+ \frac{1}{4V}
\left( \frac{2 dy}{y}-\delta dv \right)
\overline{\left( \frac{2 dy}{y}-\delta dv \right)} .
\end{eqnarray}
So it becomes \cite{NOYY} K\"ahlerian.

             Let us consider a M fivebrane
which worldvolume is topologically  ${\bf R}^4 \times {\bf C}$.
Its four-dimensional part ``${\bf R}^4$'' is mapped into the
$(x^0,x^1,x^2,x^3)$-space and identified with it.
Its two-dimensional part denoted by the complex plane is supposed
to be holomorphically embedded into the above Taub-NUT space.
Let us consider the holomorphic embedding of the following type
\begin{eqnarray}
y=v+e_0,
\label{M5}
\end{eqnarray}
where ``$y$'' is the complex function of $v,b$ and $\sigma$
given in eq.(\ref{y}).
As we shall show soon,
eq.(\ref{M5}) describes a configuration
consisting of the fivebrane sitting to the right of the sixbrane
in the Taub-NUT space.
Using expression (\ref{y}) one can separate eq.(\ref{M5})
into radial and angular parts. The radial part acquires the form
\begin{eqnarray}
e^{-b}=b+b_0
+\sqrt{(b+b_0)^2+|v+e_0|^2},
\label{real part}
\end{eqnarray}
while the angular part becomes
\begin{eqnarray}
e^{-i \frac{\sigma}{2}}=
\frac{v+e_0}{|v+e_0|}.
\label{imaginary part}
\end{eqnarray}
Notice that these two equations determine respectively $b$ and
$\sigma$ as the functions of $v$ and $\bar{v}$.
Let us denote them by $b(v)$ and $\sigma(v)$ for brevity.

        We shall first examine the solution $\sigma(v)$ for
eq.(\ref{imaginary part}). It actually describes a vortex at
$v=-e_0$. This vortex corresponds to an intersection of
the Dirac string of the $U(1)$ gauge potential $A$ and
the fivebrane
\footnote{It is the intersection in the $(v,b)$-space.
To discuss their intersection in the Taub-NUT space
we should replace the Dirac string by the non-compact two-cycle
projected to the Dirac string. But, even in the Taub-NUT space,
formulas described in terms of differential forms
such as eq.(\ref{Dirac string})
do not change.}.
It can be explained by the following argument :
Consider the sigma model action of the fivebrane.
Its bosonic part will include the term,
$\int d^6 \xi \sqrt{|h|} h^{ij} \partial_i X^M \partial_j X^N G_{MN}(X)$,
where $\xi^i (0 \leq i \leq 5)$ is the worldvolume coordinates and
$h_{ij}$ is an auxiliary metric on the worldvolume.
By rewriting \cite{Inami} the eleven-dimensional metric $G_{MN}$
under the $S^1$-compactification in terms of the ten-dimensional
quantities $(g_{mn},A_m,e^{\phi})$ $(0 \leq m,n \leq 9)$,
it can be seen that the above bosonic part induces the term,
$\int d^6 \xi
\sqrt{|h|}e^{\frac{4}{3}\phi}h^{ij}\nabla_iX^{10}\nabla_jX^{10}$,
where $\nabla_iX^{10} \equiv \partial_iX^{10}+\partial_iX^mA_m(X)$.
This gives an interaction of $X^{10}(\sim \sigma)$
with the $U(1)$ gauge potential.
Through this interaction the Dirac string will
make a vortex (of $\sigma$) on the fivebrane worldvolume.

          It should be also noticed that,
since the sixbranes are the magnetic sources of the $U(1)$ gauge
potential, the number of the vortices made by the intersection
between the Dirac strings (or their non-compact two-cycles)
and the fivebranes define
\cite{B-T} the linking number \cite{H-W} among them.
As for the configurations determined by (\ref{M5}) it is an
invariant.

             Next let us pay attention to the solution $b(v)$ for
eq.(\ref{real part}). The above argument makes the fivebrane
intersect with the Dirac string which we have just fixed
in eq.(\ref{Dirac string}). Therefore it implies the inequality
\begin{eqnarray}
\left.b(v)\right|_{v=-e_0} > -b_0,
\label{S-M intersection}
\end{eqnarray}
which means that the fivebrane is located on the ``right'' of
the sixbrane in the Taub-NUT space.

       At this stage it might be convenient to describe
a process of brane creation  in a very heuristic manner
\footnote{A similar observation, but in a flat background,
was also made in \cite{S-S}. We thank the authors of \cite{S-S}
for letting us know about it.}.
For this purpose we shall consider the holomorphic embedding of the fivebrane
when the position of the sixbrane $\vec{r}_0=(-e_0,-b_0)$ goes to infinity.
Let us examine the situations, $-b_0 \rightarrow \pm \infty$.
We first remark that, in the region
$|v+e_0| \ll |b+b_0|$,
the R.H.S of eq. (\ref{real part}) behaves approximately as
\begin{eqnarray}
b+b_0 +\sqrt{(b+b_0)^2+|v+e_0|^2}
\sim
(b+b_0+|b+b_0|)+\frac{|v+e_0|^2}{2|b+b_0|}.
\label{estimate1}
\end{eqnarray}
Owing to this estimate,
taking care of the absolute value in it,
eq.(\ref{real part}) will acquire the following form
at the limit $-b_0 \rightarrow +\infty$
\begin{eqnarray}
e^{-\frac{b}{2}}= const. |v+e_0|,
\label{estimate2}
\end{eqnarray}
while it will become at the limit
$-b_0 \rightarrow -\infty$
\begin{eqnarray}
e^{-\frac{b}{2}}= const.
\label{estimate3}
\end{eqnarray}
Notice that the Taub-NUT space itself can be regarded
as the flat space $Q_0 = {\bf R}^3 \times S^1$ in these limits.
So, according to the discussion given in \cite{Witten},
the first case describes, in IIA picture,
a configuration consisting of a NS fivebrane and a semi-infinite
D fourbrane sticking to the NS fivebrane from the right
while the second case describes a configuration of a NS fivebrane.
This will mean that the sixbrane running from
$x^6=-\infty$ to $x^6=+\infty$ creates a D fourbrane
by crossing the fivebrane.
This may give an heuristic explanation for the brane creation
via M theory. In spite of its intuitiveness it is still too
naive to justify. So let us provide a rigorous treatment.

\subsubsection*{Detail on M fivebrane}

         Without loss of generality we can set the position of the
sixbrane $\vec{r}_0=0$ by a parallel shift of the coordinates.
In such coordinates eq.(\ref{real part}) acquires the form
\begin{eqnarray}
e^{-b}=A \left(b + \sqrt{b^2+|v|^2}\right),
\label{eq for b}
\end{eqnarray}
where $A \equiv e^{-b_0}$. Since it does not depend on
the phase of $v$ the solution for this equation is a function
of $|v|$ and $b_0$.
Let us denote it by $b(|v|;b_0)$.
Notice that relation (\ref{S-M intersection}) implies
\begin{eqnarray}
\beta(b_0) \equiv b(0;b_0) > 0 .
\label{estimate on beta}
\end{eqnarray}
Let us start by describing how the solution depends on
$|v|$ and $b_0$. It can be given by partially-differentiating
eq.(\ref{eq for b}) with respect to $|v|$ and $b_0$.
After a little calculation we obtain
\begin{eqnarray}
\frac{\partial b(|v|;b_0)}{\partial |v|}
&=& -\frac{1}{1+\Delta(|v|;b_0)}
\frac{-b(|v|;b_0)+\Delta(|v|;b_0)}{|v|} ,
\label{v-dependence}
\\
\frac{\partial b(|v|;b_0)}{\partial b_0}
&=& \frac{\Delta(|v|;b_0)}{1+\Delta(|v|;b_0)} ,
\label{b0-dependence}
\end{eqnarray}
where we introduce $\Delta(|v|;b_0)$ by
\begin{eqnarray}
\Delta(|v|;b_0)\equiv
\sqrt{b(|v|;b_0)^2+|v|^2}.
\label{Delta}
\end{eqnarray}
Since it holds
$\partial b(|v|;b_0)/\partial |v| <0$ for $|v|>0$ and
$\partial b(|v|;b_0)/\partial b_0 >0$,
it means that $b(|v|;b_0)$ is monotonically decreasing
with respect to $|v|$ while it is monotonically increasing
with respect to $b_0$.

         Let us consider the behavior of $\beta(b_0)=b(0;b_0)$.
Eq.(\ref{b0-dependence}) implies the following 1st order differential
equation for $\beta$
\begin{eqnarray}
\frac{d \beta (b_0)}{d b_0}=
\frac{\beta(b_0)}{1+\beta(b_0)},
\label{eq for beta}
\end{eqnarray}
which can be integrated out. One can find
$e^{b_0}=C\beta(b_0)e^{\beta(b_0)}$ where
$C$ is a positive constant. Owing to this form there exists
a critical value $b_0^*$ such that
$b_0 \leq b_0^*$ implies $\beta(b_0)\geq b_0$
and vice versa. In such a case,
since it has been pointed out that $b(|v|;b_0)$ is
monotonically decreasing with respect to $|v|$,
there exists $r(b_0)$ which satisfies
$b(r(b_0);b_0)=b_0$. An explicit form of $r(b_0)$
can be obtainable from eq.(\ref{eq for b}) by inserting
$(|v|,b)=(r(b_0),b_0)$ into it
\begin{eqnarray}
r(b_0)=\sqrt{1-2b_0}.
\label{r}
\end{eqnarray}
Since $r(b_0)$ is, by definition, a  non-negative quantity
and it must be zero at the critical
value of $b_0$, we can find out $b_0^*$ as
\begin{eqnarray}
b_0^*=1/2.
\label{b0*}
\end{eqnarray}
Notice that it holds $\beta(b_0^*)=b_0^*$.
Inserting this value into the integrated form
$e^{b_0}=C\beta(b_0)e^{\beta(b_0)}$, we obtain $C=2$.
Hence $\beta(b_0)$ is determined by the relation
\begin{eqnarray}
2\beta(b_0)e^{\beta(b_0)}=e^{b_0} .
\label{beta}
\end{eqnarray}

      Finally we shall consider the intersection between
the fivebrane and $b=0$ plane in the Taub-NUT space. It is
a circle with a definite radius. The radius $u(b_0)$ can be read
from eq.(\ref{eq for b}). It is given by
\begin{eqnarray}
u(b_0)=e^{b_0}.
\label{u}
\end{eqnarray}

\begin{figure}[t]
\epsfysize=8cm \centerline{\epsfbox{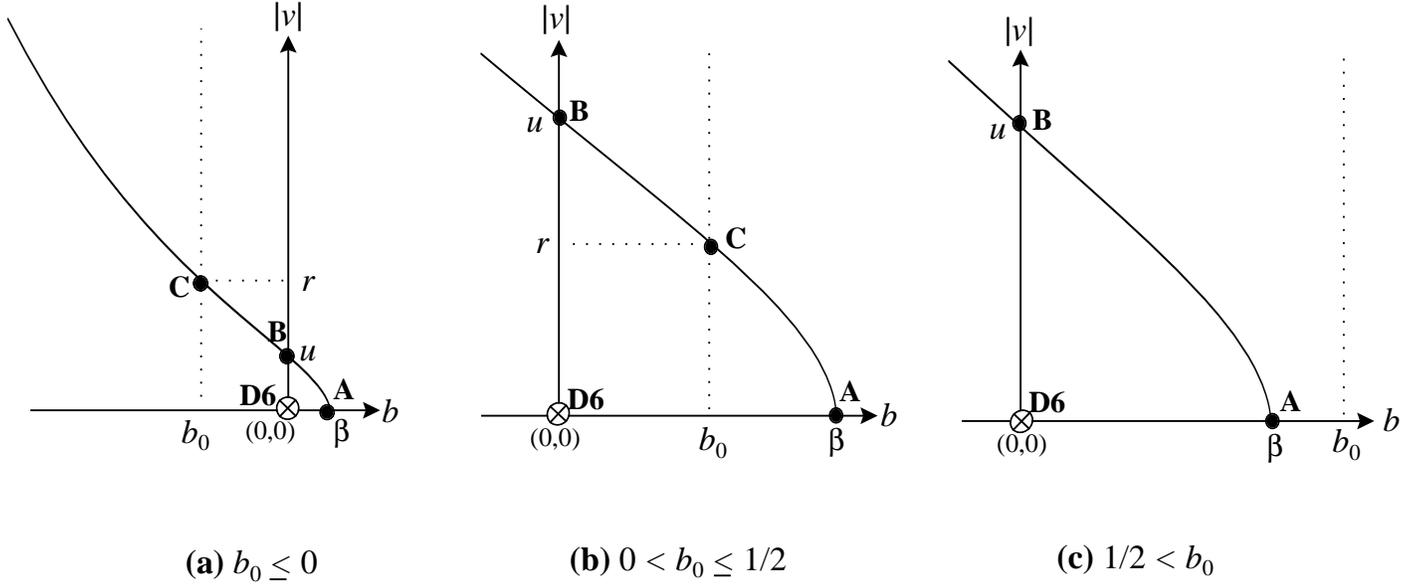}}
\caption{\small
Shapes of the fivebrane in the $(|v|,b)$-plane
}
\label{three cases}
\end{figure}

      Now, gathering all these data about some characteristic points of
the embedded two-plane, it is possible to draw the shape of the fivebrane
in the Taub-NUT space.
There appear three cases depending on the values of $b_0$ :
$b_0 \leq 0$, $0< b_0 \leq 1/2$ and $b_0 > 1/2$.
In Fig.\ref{three cases} these three cases are depicted in the $(|v|,b)$-plane.
Three characteristic points are denoted by A, B and C.
Their coordinates in the $(|v|,b)$-plane are respectively
$(0,\beta(b_0)),(u(b_0),0)$ and $(r(b_0),b_0)$.

\subsubsection*{IIA limit}

Generically the IIA limit of the above configurations
will be achieved by taking the limit $R \rightarrow 0$
with fixing the value of $Rb_0/2$
\footnote{The factor $1/2$ is our convention.} \cite{IIA limit}.
(Therefore $b_0$ goes to infinity.)
Let us denote this value by $x^6_0$.
We first examine the case of $b_0 \leq 0$.
In this case one should set $x^6_0 \leq 0$ because of the
relation $x^6_0=Rb_0/2$.
And, to fix $x^6_0$ at a finite negative value, one must take
$b_0 \rightarrow -\infty$.
Let us consider the IIA limits of the points A, B and C.
The point A will become the point given by the coordinates
$(|x^4+ix^5|,x^6)=(0,\mbox{lim}_{R \rightarrow 0}R\beta(b_0)/2)$.
What is the value of $\mbox{lim}_{R \rightarrow 0}R\beta(b_0)$?
We shall first rewrite eq.(\ref{beta}) into the form
\begin{eqnarray}
R \beta(b_0)+R \ln R\beta(b_0)-R \ln 2R=2x_0^6.
\label{IIA beta 1}
\end{eqnarray}
Notice that, owing to eq.(\ref{estimate on beta}),
$R \beta(b_0)$ is a positive quantity at a non-zero value of $R$.
Therefore, in order to fix the R.H.S. of the equation at a negative
value, $R \ln R \beta(b_0)$ in the L.H.S. must become a negative quantity
as $R$ goes to zero.
This implies $\mbox{lim}_{R \rightarrow 0}R \beta(b_0)=0$.
So the point A approaches to the origin.
Next we examine the point B.
It becomes the point given by the coordinates
$(|x^4+ix^5|,x^6)=(\mbox{lim}_{R \rightarrow 0}R u(b_0)/2,0)$.
Using the expression given in (\ref{u}) one can also see
$\mbox{lim}_{R \rightarrow 0} R u(b_0)=0$.
Hence, the point B also approaches to the origin under the IIA limit.
As regards the point C,
It becomes the point given by the coordinates
$(|x^4+ix^5|,x^6)=(\mbox{lim}_{R \rightarrow 0}Rr(b_0)/2,x^6_0)$.
By the expression given in (\ref{r}) one can see
$\mbox{lim}_{R \rightarrow 0}R r(b_0)=0$.
So, the point C approaches to $(0,x_0^6)$.

\begin{figure}[t]
\epsfysize=8cm \centerline{\epsfbox{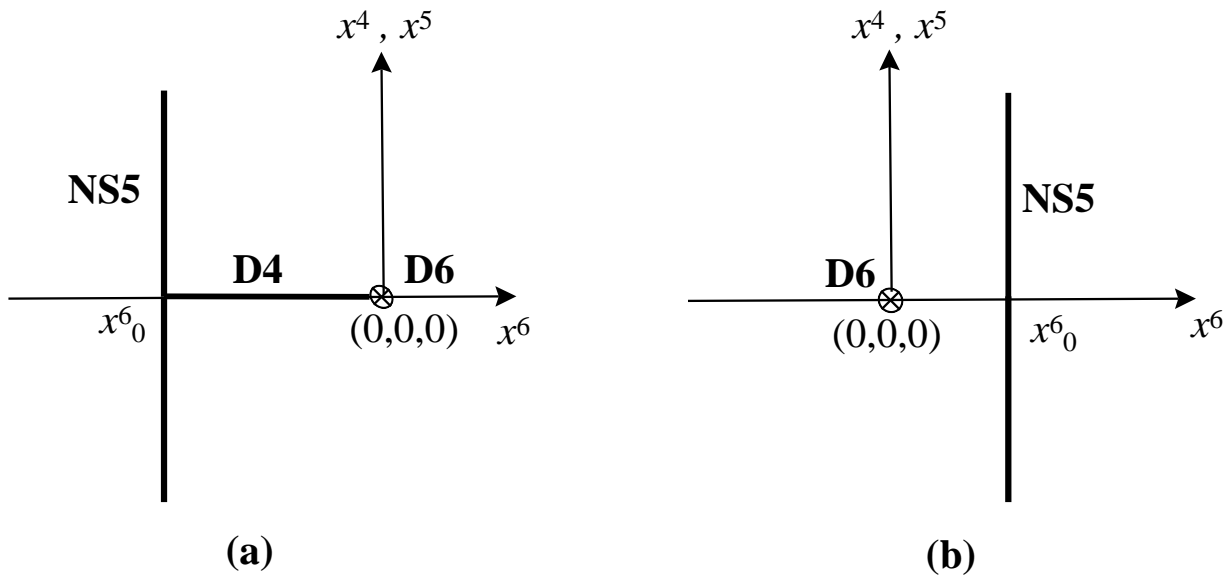}}
\caption{\small
IIA limits of M-theory brane configurations depicted in Fig.1  :
(a) D fourbrane is suspended between NS fivebrane and D sixbrane.
(b) There are only NS fivebrane and D sixbrane.
}
\label{iialimit}
\end{figure}

         The segment between the points A and C
on the fivebrane in the $(|v|,b)$-plane
reduces, in the IIA limit, to a line segment $(0,x^6)$
where $x^6_0 \leq x^6 \leq 0$.
Notice that this line segment in the $(|x^4+ix^5|,x^6)$-plane
is still one-dimensional in the three-dimensional
$(x^4,x^5,x^6)$-space
since $|x^4+ix^5|=0$ implies $x^4=x^5=0$.
How about the other part of the fivebrane?
To obtain its IIA limit
let us rewrite eq.(\ref{v-dependence}) in terms of $x^6$ and $|x^4+ix^5|$.
By introducing the quantity $t = b/|v|$ $(=x^6/|x^4+ix^5|)$ it becomes
\begin{eqnarray}
\frac{\partial x^6}{\partial |x^4+ix^5|}&=&
\mbox{lim}_{R \rightarrow 0}
R \frac{t-\sqrt{1+t^2}}{R+|x^4+ix^5|\sqrt{1+t^2}}
\nonumber \\
&=& 0.
\end{eqnarray}
So, this part of the fivebrane approaches to the line
$(|x^4+ix^5|,x^6_0)$. Due to the rotational symmetry around
the $x^6$-axis this line corresponds to a two-dimensional
plane in the $(x^4,x^5,x^6)$-space.
Therefore the IIA limit of the configuration with $b_0 \leq 0$
describes the configuration of a NS fivebrane,
a D fourbrane and a D sixbrane.
The NS fivebrane is located at $(x^6,x^7,x^8,x^9)=(x^6_0,0,0,0)$
while the D fourbrane with worldvolume $(x^0,x^1,x^2,x^3,x^6)$
is suspended, in the $(x^4,x^5,x^6)$-space,
between the point $(x^4,x^5,x^6)=(0,0,x^6_0)$
on the NS fivebrane and the D sixbrane at $(x^4,x^5,x^6)=(0,0,0)$.
(Fig.\ref{iialimit} (a))

      Next we shall consider the case of $0 < b_0 \leq 1/2$.
In this case we must set $x^6_0>0$. So, $b_0 \rightarrow +\infty$.
The IIA limit of the point A is given by the coordinates
$(|x^4+ix^5|,x^6)=(0,\mbox{lim}_{R \rightarrow 0}R \beta(b_0)/2)$.
Let us show $\mbox{lim}_{R \rightarrow 0}R \beta(b_0)=x_0^6$.
Rewrite eq.(\ref{beta}) into the form
\begin{eqnarray}
R \left( \beta(b_0)+\ln 2\beta(b_0) \right)=2x_0^6.
\label{IIA beta 2}
\end{eqnarray}
In order to fix the R.H.S. of the equation at a positive value,
$\beta +\ln 2\beta$ in the L.H.S. must approach to $+\infty$
as $R$ goes to zero. In particular it holds $\beta \rightarrow +\infty$.
Hence $\mbox{lim}_{R \rightarrow 0} \ln \beta /\beta =0$.
Regarding the L.H.S. of eq.(\ref{IIA beta 2}) as
$R \beta(b_0)(1+ \ln 2\beta(b_0)/\beta(b_0))$ and considering
its IIA limit, we obtain
$\mbox{lim}_{R \rightarrow 0}$ $R \beta(b_0)/2$ $=x_0^6$.
Therefore the point A approaches to $(0,x^6_0)$ in the IIA limit.
As for the point C, it becomes the point given by the coordinates
$(\mbox{lim}_{R \rightarrow 0}R r(b_0)/2, x^6_0)$.
With the same reason as in the previous case
it is $(0,x^6_0)$ in the IIA limit.
Finally the point B becomes
the point given by the coordinates
$(\mbox{lim}_{R \rightarrow 0}Ru(b_0)/2,0)$.
Since $R u(b_0)= 2R \beta(b_0)e^{R\beta(b_0)/R}$
behaves as $2x^6_0e^{x^6_0/R}$ at a small value of $R$
and goes to the infinity after the IIA limit,
it holds $\mbox{lim}_{R \rightarrow 0}Ru(b_0)=+\infty$.
To summarize, the IIA limit of the configuration with
$0 < b_0 \leq 1/2$  describes the configuration of
a NS fivebrane and a D sixbrane. The NS fivebrane
is located at $(x^6,x^7,x^8,x^9)=(x^6_0,0,0,0)$
and the D sixbrane is at $(x^4,x^5,x^6)=(0,0,0)$.
There appears no D fourbrane.
(Fig.\ref{iialimit} (b))

            Finally the study for the case of $b_0 > 1/2$
is remained. But it is easy to see that this case reduces to
that of $0 < b_0 \leq 1/2$. In particular
the IIA limit coincides with the previous case.

\section*{Acknowledgments}

T.N. is supported in part by
Grant-in-Aid for Scientific Research 08304001.
K.O. and Y.Y. are supported in part by the JSPS
Research Fellowships.

\end{document}